\documentclass[prb,preprint,showpacs]{revtex4}
\usepackage{epsfig}
\usepackage{amsmath}

\begin{document}
\title{Quantum dot defined
in two-dimensional electron gas at n-AlGaAs/GaAs heterojunction:
simulation of electrostatic potential and charging properties}
\author{S. Bednarek}
\author{K. Lis}
\author{B. Szafran}
\affiliation{Faculty of Physics and Applied Computer Science, AGH
University of Science and Technology, al. Mickiewicza 30, 30-059
Krak\'ow, Poland}

\date{\today}

\begin{abstract}
We present a self-consistent Schroedinger-Poisson scheme for
simulation of electrostatic quantum dots defined in gated
two-dimensional electron gas formed at n-AlGaAs/GaAs heterojunction.
The computational method is applied to a quantitative description of
transport properties studied experimentally by Elzermann et al.
[Appl. Phys. Lett. {\bf 84}, 4617 (2004)]. The three-dimensional
model describes the electrostatics of the entire device with a
quantum dot that changes shape and floats inside a gated region when
the applied voltages are varied. Our approach accounts for the metal
electrodes of arbitrary geometry and configuration, includes
magnetic field applied perpendicular to the growth direction,
electron-electron correlation in the confined electron system and
its interaction with the electron reservoir surrounding the quantum
dot. We calculate the electric field, the space charge distribution
as well as energies and wave functions of confined electrons to
describe opening of two transport channels between the reservoir and
the confined charge puddle. We determine the voltages for charging
the dot with up to 4 electrons.
 The results are in a qualitative and
quantitative agreement with the experimental data.
\end{abstract}
 \maketitle

\section{Introduction}
A quest for a nanodevice that would store a quantum bit in the spin
\cite{1,2} of a confined electron and allow for its manipulation is
a main factor that motivates the research on gated electrostatic
quantum dots. Realization of a spin quantum gate requires
application of multiple quantum dots coupled in way that can be
controlled during the device operation. First electrostatic quantum
dots were formed in a vertical configuration \cite{3,4,5,6,7} of a
semiconductor heterostructure containing a single or multiple
quantum well surrounded by a single gate creating the lateral
confinement. The strength of the tunnel coupling between the
vertical dot and electron reservoirs depends on the applied barrier
thickness and composition, and therefore it is fixed for each
device. Similarly, for vertical artificial molecules the interdot
coupling is defined at the production stage.

A full control of the interdot coupling is possible in quantum dots
formed in gated two-dimensional electron
gas.\cite{8,9,10,11,12,13,14} These structures are produced by
deposition of multiple gates on top of the n-AlGaAs/GaAs structure
containing a two-dimensional electron gas at the heterojunction. The
system of gates is designed to locally deplete the electron gas in
order to tailor a quantum dot of the surrounding electron reservoir.
The voltages applied to the multiple gates define the confinement
potential for the trapped electrons and control the tunnel barrier
between the dot and the reservoir. Thus, the coupling of the
confined artificial atom to the environment can be intentionally
tuned from an open to a closed dot regime.  For nearly open dots the
Kondo effect and co-tunneling phenomena are observed.\cite{kondo}
Coupled dots with voltage tunable interdot barriers are also
produced.\cite{13,14,15}
 Quantum dots formed in the gated two-dimensional electron gas
are used for investigation of spin dependent transport\cite{spindt}
and confined spin relaxation.\cite{spinrl}  A capacitive coupling
between the dot and a quantum point contact defined in the same
structure\cite{8,9,10,11,12} is used to probe the confined states by
the conductance measurements. The purpose of the present paper is to
provide a theoretical description of a quantum dot formed in the
gated two-dimensional electron gas. We focus our attention on the
nanodevice that is probed by the quantum point contact as described
in Refs. [\cite{9}] and [\cite{10}].

The electrostatic confinement potential in vertical quantum dots was
thoroughly studied in Refs.[\cite{16,17,18,19,20,21,22,23}]. Less
attention was paid to dots based on the gated two-dimensional
electron gas. In particular, a theory for a double planar dot
\cite{14} was provided in Refs. [\cite{24}] and [\cite{u}]. Ref.
[\cite{15}] describes a structure of a triple quantum dot.
Theoretical modeling of planar dots is for several reasons more
difficult than modeling of the vertical structures. In vertical dots
the electrons are confined inside a relatively deep quantum well
with the lateral confinement strength controlled by a {\it single}
gate. In planar structures the confinement is entirely due to the
voltages that are applied to multiple gates and the formed potential
cavity is typically shallow. Therefore, both the confinement
potential and the few-electron eigenproblem have to be calculated
with a special care. Moreover, as we show below, the gates not only
fix the strength of the dot confinement but also change the shape of
confined charge island which floats with the voltages inside the
gated area. Interaction of the confined electrons with the
reservoirs is also more complex due to variation of the depleted gas
region with the applied potentials. The vertical
dots\cite{16,17,18,19,20,21,22} often have circular
symmetry,\cite{uwa2} which is not the case for the planar structure
discussed below.

\section{Theory}

\subsection{Model of the structure}

The nanodevice \cite{9,10} which we aim to describe is constructed
on a basis of a planar semiconductor heterostructure of
GaAs/n-Al$_{0.3}$Ga$_{0.7}$As in which the two-dimensional electron
gas is created at the GaAs side of the junction. A cross section of
the layer structure is presented in Fig. 1. On the substrate side
there is a 1500 nm thick layer of undoped GaAs with a blocking
AlGaAs barrier deposited on top. Lower part of the barrier (20 nm
thick) is undoped and the upper (65 nm) is heavily doped with
donors. On top of the barrier there is a 5 nm thin layer of n-doped
GaAs. The donor states in AlGaAs stay 200 meV above the conduction
band minimum of GaAs. Therefore, the electrons pass to the GaAs
layer but stay localized under the barrier due to the Coulomb
attraction by the charge of ionized donors.

\begin{figure}[ht!]
\centerline{ \hbox{\epsfysize=45mm
                \epsfbox[181 428 458 668] {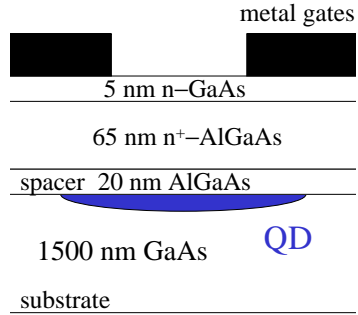}\hfill}}
\caption{(color online) Structure of layers used for formation of a
gated quantum dot in the two-dimensional electron gas (according to
Ref. [\cite{9}]).
 \label{pot}}
\end{figure}

In our model we assume the shape of gates deposited on top of the
structure (see Fig. 2) according to Refs. [\cite{9,10}]. A negative
voltage applied to the electrodes depletes the electron gas
underneath and forms a lateral confinement potential in the center
of the gated area. For properly adjusted voltages a few electrons
stay localized in the center of the structure forming an artificial
atom. The electrons are confined in the vertical direction by the
barrier formed due to the GaAs/AlGaAs conduction band offset $U_b$.
The electrostatic potential $\phi_\mathrm{elst}({\bf r})$ is
responsible for the lateral confinement as well as for the potential
that closes the dot from the substrate side.
 The quantum dot confinement potential is therefore
expressed by
\begin{equation}
U_\mathrm{conf}({\bf r})=U_b(z)-|e|\phi_\mathrm{elst}({\bf r}).
\label{eq1}
\end{equation}
 $U_b$ equals zero in GaAs layer and the conduction band offset
in the AlGaAs barrier. The electrons are additionally confined by an
in-plane magnetic field of 10 T applied parallel to the surface of
the layer structure.\cite{8,9} Application of a strong magnetic
field within the plane of confinement\cite{Molinari} was previously
discussed in context of tuning (reduction) of the electron tunnel
coupling for vertical artificial molecules. To the best of our
knowledge the Schroedinger-Poisson problem for the two dimensional
electron gas with an in-plane orientation of the magnetic field was
never solved previously.

\subsection{Sources of electrostatic potential}

The total electrostatic potential $\phi_\mathrm{tot}({\bf r})$ is
influenced by voltages applied between the substrate and metal
electrodes on top of the structure as well as by the charge
distribution inside the device. The charge density
$\rho_\mathrm{tot}({\bf r})$ is a sum of three contributions that
have different nature and distribution
\begin{equation}
\rho_\mathrm{tot}({\bf r})=\rho_\mathrm{eqd}({\bf
r})+\rho_\mathrm{d}({\bf r})+\rho_\mathrm{el}({\bf r}).
\end{equation}
The first contribution $\rho_\mathrm{eqd}({\bf r})$ is the
distribution of the charge confined in the quantum dot that is found
by the solution of a few-electron quantum eigenproblem that is in
the present work obtained by the configuration interaction method.
The second contribution  $\rho_\mathrm{d}({\bf r})$ is the ionized
donor space charge in the AlGaAs barrier, and the third
$\rho_\mathrm{el}({\bf r})$ is the charge density of the electron
gas.

\begin{figure}[ht!]
\centerline{ \hbox{\epsfysize=65mm
                \epsfbox[140 572 355 793] {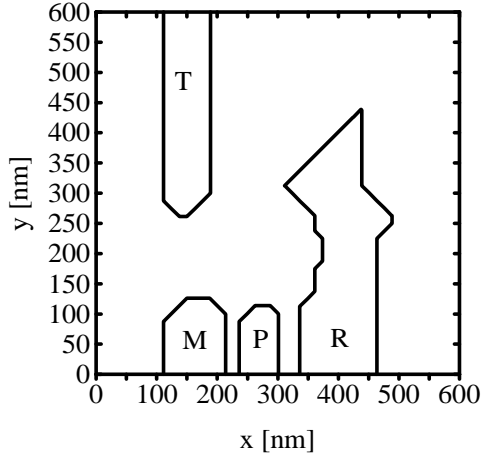}\hfill}}
\caption{Geometry, size and position of the metal gates deposited on
top of the semiconductor heterostructure (according to Ref.
[\cite{9}]).
 \label{gates}}
\end{figure}

According to the superposition principle the total electrostatic
potential $\phi_\mathrm{tot}({\bf r})$ can be expressed as a sum of
contributions of all the charge densities. We separate the potential
due to the confined electrons $\phi_\mathrm{eqd}$ of the total
potential. In this way we obtain a component of the potential
$\phi_\mathrm{elst}$ that enters the
 formula (\ref{eq1}) for the confinement potential of the
electrostatic dot
\begin{equation}
\phi_\mathrm{tot}({\bf r})=\phi_\mathrm{eqd}({\bf
r})+\phi_\mathrm{elst}({\bf r}).
\end{equation}
Potential of the dot-confined charge is calculated directly from the
Coulomb law
\begin{equation}
\phi_\mathrm{eqd}({\bf r})=\frac{1}{4\pi\epsilon\epsilon_0} \int
\frac{\rho_\mathrm{eqd}({\bf r}')}{|{\bf r}-{\bf r}'|}d{\bf r}'.
\end{equation}
Potential $\phi_\mathrm{elst}$ of all the other sources is found
from the Poisson equation
\begin{equation}
\nabla^2 \phi_\mathrm{elst}({\bf r})=-\frac{\rho({\bf r})}{\epsilon
\epsilon_0}.\label{poisson}
\end{equation}
The proper boundary conditions are naturally given for the total
potential $\phi_\mathrm{tot}({\bf r})$. The boundary conditions for
the Poisson equation (\ref{poisson}) are obtained from boundary
conditions that are known for the total potential
$\phi_\mathrm{tot}({\bf r})$ by a simple subtraction
\begin{equation}\phi_\mathrm{elst}({\bf r})=\phi_\mathrm{tot}({\bf
r})-\phi_\mathrm{eqd}({\bf r}). \label{pipa}
\end{equation}
Charge density $\rho({\bf r})$ in Eq. (\ref{poisson}) is a sum of
the charge densities of ionized donors and of the electron gas
\begin{equation}
\rho({\bf r})=\rho_\mathrm{d}({\bf r})+\rho_\mathrm{el}({\bf r}).
\label{ro}
\end{equation}
The donors are ionized in the spatial region where the {\it total}
electron potential energy calculated with respect to GaAs layer is
larger than the donor binding energy. For the donor energy level
taken as the reference energy we obtain the ionization condition
\begin{equation}
\rho_\mathrm{d}({\bf r})=\left\{\begin{array}{lcr} 0 & \mathrm{ for
} & -|e|\phi_\mathrm{tot}({\bf r})\leq E^D \\
 |e|n_\mathrm{D}({\bf r}) & \mathrm{ for } & -|e|\phi_\mathrm{tot}({\bf
 r})>E^D
 \end{array}\right., \label{ic}
\end{equation}
where $n_\mathrm{D}({\bf r})$ is the density of donor impurities,
and $E^D$ is the donor binding energy. Assumption of a homogenous
(continuous) donor distribution  is justified by the presence of an
undoped AlGaAs buffer (see Fig. 1).

Calculation of $\rho_\mathrm{el}({\bf r})$ -- the second charge
density entering Eq. (\ref{ro}) that corresponds to the electron gas
confined at the heterojunction -- is a nontrivial task. It can be
exactly evaluated only in the asymptotic region, i.e. at a large
distance of the electrodes. In the neighborhood of the electrodes
one has to introduce an approximate treatment (see II.D).

\subsection{Potential and charge distribution in the asymptotic region}

At a large distance of the electrodes the electric field is parallel
to the growth direction ($z$) and the electrostatic potential does
not depend on the other two coordinates. The potential distribution
results from an equilibrium between the ionized donors in the AlGaAs
layer and the electron gas confined below. We choose the $y$ axis as
parallel to the external magnetic field. We adapt the Landau gauge
 \begin{equation}{\bf A} ({\bf
r})=(-Bz,0,0)\end{equation} which leads to the Hamiltonian of an
electron confined at the heterojunction
\begin{eqnarray}
H({\bf r})&=&-\frac{\hbar^2}{2m}\left(\frac{\partial ^2}{\partial
x^2}+\frac{\partial ^2}{\partial y^2}+\frac{\partial ^2}{\partial
z^2}\right)+i\hbar \omega_cz\frac{\partial }{\partial x}\nonumber
\\&&+
\frac{m}{2}\omega_c^2z^2+U_b(z)-|e|\phi_\mathrm{elst}(z),
\label{ham}
\end{eqnarray}
where $\omega_c=\frac{|e|B}{m}$ and $m$ is the electron band mass.
Since this Hamiltonian commutes with momentum in both $x$ and $y$
directions its eigenfunctions are expected to be of form
\begin{equation}
\Psi^\mathrm{as}({\bf r})=C\exp(iqx)\exp(iky)\phi^\mathrm{as}(z).
\label{eigenf}
\end{equation}
Eigenequation for $\phi^\mathrm{as}(z)$ is obtained by substitution
of Eq. (\ref{eigenf}) into (\ref{ham})
\begin{equation}
H(z)\phi^\mathrm{as}_{nq}(z)=\varepsilon_{nq}\phi_{nq}^\mathrm{as}(z),
\label{se}
\end{equation}
where
\begin{equation}
H(z)=-\frac{\hbar^2}{2m}\frac{\partial^2}{\partial
z^2}+\frac{m}{2}\omega_c^2(z-z_0)^2+U_b(z)-|e|\phi_\mathrm{elst}^\mathrm{as}(z),
\end{equation}
with $z_0={\hbar q}/{m\omega_c}$.

Eigenvalues $\varepsilon_{nq}$ and the eigenfunctions
$\phi_{nq}^\mathrm{as}(z)$ are labeled by a natural quantum number
$n$ and depend on the wave vector $q$ in the direction perpendicular
to the magnetic field direction. Wave vector $q$ enters the
Hamiltonian operator (13) through shifted harmonic oscillator
minimum $z_0$. The total electron energy eigenvalues are given by
\begin{equation}
E_{nkq}=\varepsilon_{nq}+\frac{\hbar^2 k^2}{2m}.
\end{equation}
The electrons confined at the heterojunction have energies below the
Fermi energy.  In the present calculations the Fermi energy is taken
as the reference energy level. Only the states with $E_{nkq}<0$ can
be confined at the heterojunction. Given the $H(z)$ eigenfunctions
 one calculates the charge density of the electron gas
\begin{eqnarray} \rho^\mathrm{as}_\mathrm{el}(z)&=&-2|e|\sum_{nkq}^{E_{nkq}<0}
|\Psi^\mathrm{as}_{nkq}(x,y,z)|^2 \\
&&=-\sum_n\frac{|e|}{2\pi^2}\int_{E_{nkq}<0}
dkdq|\phi^\mathrm{as}_{nq}(z)|^2\nonumber \\
&&=-\sum_n\frac{|e|}{2\pi^2}\int_{\varepsilon_{nq}<0}dq\int_{-k_F}^{k_F}dk
|\phi^\mathrm{as}_{nq}(z)|^2, \nonumber
\end{eqnarray}
where $k_F=\sqrt{-\frac{2m\varepsilon_{nq}}{\hbar^2}}$. By
integration over $k$ we obtain
\begin{equation}
\rho^\mathrm{as}_\mathrm{el}(z)=-\sum_n\frac{|e|}{\pi^2}\int_{\varepsilon_{nq}<0}dq\sqrt{-\frac{2m\varepsilon_{nq}}{\hbar^2}}
|\phi^\mathrm{as}_{nq}(z)|^2. \label{cd}
\end{equation}
Apart from the electron gas another source of the electric field is
the ionized donor distribution $\rho_\mathrm{d}$ calculated
according to Eq. (\ref{ic}) in which  we identify
$\phi_\mathrm{tot}^\mathrm{as}=\phi_\mathrm{elst}^\mathrm{as}$ since
at an asymptotically large distance of the quantum dot potential
$\phi_\mathrm{eqd}$ vanishes. Finally, the electrostatic potential
is calculated from a single-dimensional Poisson equation
\begin{equation}
\frac{\partial ^2}{\partial
z^2}\phi_\mathrm{elst}^\mathrm{as}(z)=-\frac{\rho_\mathrm{el}^\mathrm{as}(z)+\rho_\mathrm{d}(z)}{\epsilon\epsilon_0}.
\label{p1}
\end{equation}
Since $\rho^\mathrm{as}_\mathrm{el}$ appearing in Eq. (17) is
calculated with the Schroedinger equation (\ref{se}), which in turn
contains the electrostatic potential, both the equations are solved
in an iteration till self-consistency is reached. Eq. (\ref{se}) is
solved below the barrier in a computational box long enough to allow
the electron charge density (\ref{cd}) to vanish before its end. The
equilibrium solution is obtained when two conditions are met. The
first one results from the charge neutrality which requires that the
number of ionized donors is equal to the number of electrons trapped
at the interface, which results in vanishing electric field at both
ends of the region where the Poisson equation (\ref{p1}) is solved.
In fact, whenever a self-consistency of Eq. (12) and (17) is
obtained the potential derivatives at both ends of the computational
box vanish. The second equilibrium condition requires that the
number of ionized donors is such that the variation of the electric
potential on the entire space charge (on both sides of the
heterojunction) equalizes the jump in the conduction band at the
GaAs/AlGaAs interface. The latter is due to the fact that the
electrons occupy all the states below the Fermi energy. This
includes both the donor states in the barrier (not all the donors
are ionized) and the states trapped below the barrier.

The surface density of the electron charge accumulated below the
barrier is obtained by integration of
$\rho^\mathrm{as}_\mathrm{el}(z)$ along the growth direction
\begin{equation}
\sigma^\mathrm{as}=\int_{-\infty}^\infty
dz\rho^\mathrm{as}_\mathrm{el}(z),
\end{equation}
with the charge neutrality condition
\begin{equation}
\sigma_D^\mathrm{as}=-\sigma^\mathrm{as},
\end{equation}
where $\sigma_D^\mathrm{as}$ is the surface density of ionized
donors. The surface densities depend on the conduction band offset.
Under the assumption of a homogenous donor distribution the
asymptotic ionization range $d$ fulfills the condition
$\sigma_D^\mathrm{as}=n_Dd$. The second equilibrium condition is
obtained for a properly chosen $d$. For a nominal composition of the
barrier Al$_x$Ga$_{1-x}$As with $x=0.27$ the barrier height is
$U_b=229$ meV. For that value one obtains the asymptotic surface
density of $\sigma^\mathrm{as}=3.5\times 10^{-11}$ cm$^{-2}$, which
is close to the nominal experimental value\cite{8} of $2.9\times
10^{-11}$ cm$^{-2}$. The deviation of the calculated density off the
nominal value may result from the neglect of the exchange
interaction in the electron gas. We decided to reduce the barrier
height to $U_b=200$ meV for which the calculated density is equal to
its nominal value.

\begin{figure}[ht!]
\centerline{ \hbox{\epsfysize=55mm
                \epsfbox[97 340 560 750] {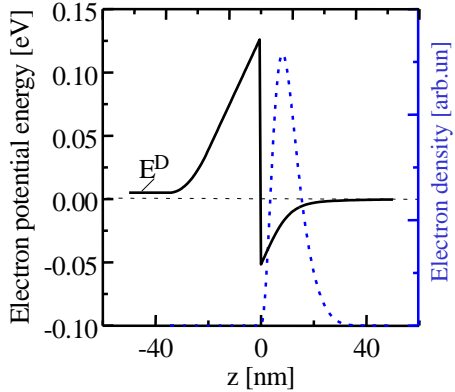}\hfill}}
\caption{Electron potential energy (solid line) and the charge
density of the electron gas (dashed curve) at the AlGaAs/GaAs
heterojunction (GaAs is at the positive side of $z$). Thin
horizontal dashed line shows the Fermi energy pinned at the donor
impurity level in AlGaAs. $E^D$ is the donor binding energy that
enters formula (8).
 \label{rers}}
\end{figure}

The potential and the electron density calculated for the asymptotic
region according to the procedure explained above is presented in
Fig. 3. From the potential dependence on $z$ we can see that the
charge neutrality condition (zero electric field at both ends of the
box) is fulfilled. We also notice that the electrostatic potentials
of both sides of the junction differ exactly by the donor binding
energy. This is because in the discussed structure the donor
impurity level in the heavily doped AlGaAs layer defines the Fermi
energy.

\subsection{Potential and charge distribution near the quantum dot}

Under the electrodes the electron potential energy is still positive
also below the AlGaAs barrier. This removes electron gas from the
region below the electrodes. The electron potential energy decreases
with the growing distance of the gates and it eventually becomes
negative in the region where the electron gas is not completely
depleted. The spatial variation of the electron density is crucial
for the shape and width of the potential barriers which separate the
quantum dot of the electron reservoirs. An account for the electron
dependence on the potential energy is taken in a following
approximate manner. We assume that the electron gas density is zero
wherever the local potential energy exceeds the Fermi energy
($E_F=0$). In region where the local potential energy is negative we
assume that the electron density is proportional to its absolute
value
\begin{equation}
\rho_\mathrm{el}(x,y,z)=\left\{\begin{array}{lcr} 0 & \mathrm{ for
}& \tau\leq 0
\\ \tau \rho^\mathrm{as}_\mathrm{el}(z)  & \mathrm{ for }& \tau >0 \end{array}\right.,
\end{equation}
 for
\begin{equation}
\tau=\frac{-|e|\phi_\mathrm{tot}(x,y,z_c)+U_b(z_c)}{-|e|\phi_\mathrm{tot}^\mathrm{as}(z_c)+U_b(z_c)},
\end{equation}
with $\tau\leq 1$, where
$\phi_\mathrm{tot}^\mathrm{as}=\phi_\mathrm{elst}^\mathrm{as}$ and
$z_c$ is the center of mass of the asymptotic electron density
\begin{equation}
z_c=\frac{1}{\sigma^\mathrm{as}}\int dz
\rho^\mathrm{as}_\mathrm{el}(z) z.
\end{equation}
The adopted formula (20) simulates the depletion of the electron gas
in the region of positive potential energy.  In the asymptotic
region
 $\tau$  tends to 1, which guarantees that the known value of the electron
density is found far away from the gates.

\subsection{Boundary conditions for Poisson equation in three dimensions}

Poisson equation \eqref{poisson} is solved in a three dimensional
rectangular region which contains the quantum dot and a sufficiently
large part of the electrodes. A standard test for the acceptable
size of the box consists in performing the calculation in function
of the box dimensions. We find that the results eventually saturate
for a rectangular box of side lengths $L_x = L_y = 600$ nm and  $L_z
= 400$ nm. The position of the adopted box with respect to the
electrodes is presented in Fig. 2. In the growth direction ($z$) the
box covers 200 nm on both sides of the AlGaAs/GaAs heterojunction.

A unique solution of the Poisson equation is obtained for boundary
condition given on the surface of the computational box and on the
metal electrodes that are inside the integration region. The
boundary conditions are naturally given for $\phi_\mathrm{tot}$. We
calculate the conditions for $\phi_\mathrm{elst}$ according to Eq.
\eqref{pipa}. On the electrodes the total potential is constant and
determined by applied voltages. On the surface of the electrodes we
assume Dirichlet boundary conditions
\begin{equation}
\phi_\mathrm{tot}=U_X+U_{S},
\end{equation}
where $X=P,T,M$ and $R$ enumerates the electrodes (see Fig. 2),
$U_X$ is the applied voltage and $U_{S}$ is the Schottky potential
which at the metal/GaAs contact is $U_{S}=-0.65$ V. On the lateral
sides of the computational box we apply Neumann boundary conditions
\begin{equation}
{\bf n}\cdot \nabla \phi_\mathrm{tot}=0,
\end{equation}
where ${\bf n}$ is the vector normal to the surface. The Neumann
conditions are consistent with the Gauss law for charge neutrality
of the computational box content. On the vertical walls of the box
(parallel to the growth direction $z$) this boundary condition is
equivalent to the assumption that the electric field is
perpendicular to the heterojunction, which agrees with the boundary
condition used in the asymptotic region.

\subsection{Electrons confined in the quantum dot}

The potential minimum that is found in the central region between
the electrodes traps several electrons provided that the applied
voltages are not too negative. Hamiltonian for the system of $N$
electrons with Landau gauge writes
\begin{eqnarray}
H_N&=&\sum_{i=1}^N \left(-\frac{\hbar^2}{2m}\nabla_i^2
+i\hbar\omega_c\left(z_i-z_o\right)\frac{\partial }{\partial
x_i}\nonumber \right.
\\ && \left. +\frac{m}{2}\omega_c^2(z_i-z_o)^2+U_b(z)-|e|\phi_\mathrm{elst}({\bf
r}_i)\right)\nonumber \\ &&
+\sum_{j=1}^N\sum_{i>j}^N\frac{e^2}{4\pi\epsilon\epsilon_0 |{\bf
r}_i-{\bf r}_j|}.
\end{eqnarray}
The eigenproblem $H_N\Psi_N=E_N\Psi_N$ is solved with a
configuration interaction approach in the basis of Slater
determinants built of single-electron wave functions that are
calculated with a finite-difference technique on a three-dimensional
mesh. The convergence for the ground state energy of 4 electrons is
achieved for the basis containing all the Slater determinants that
can be constructed of the wave functions of 10 lowest-energy
single-electron levels.

The parameter $z_o$ introduced by the gauge transformation is taken
to minimize the total energy of $N$ electrons. In our calculations
we adopt $z_o\simeq 12$ nm which coincides with the center of the
density of the dot-confined electrons \cite{uwaga} (see also Fig.
2).

The few-electron wave function is used to evaluate the confined
charge density
\begin{equation}
\rho_\mathrm{eqd}({\bf r}_1)=-|e|\int d{\bf r}_2d{\bf r}_3\dots
d{\bf r}_N|\Psi_N({\bf r}_1,{\bf r}_2,{\bf r}_3,\dots,{\bf r}_N)|^2.
\end{equation}

The ground-state energies for $N$ and $N-1$ electrons determine the
electrochemical potential of the $N$-electron quantum dot
\begin{equation}\mu_N=E_N-E_{N-1}.\end{equation} The dot is filled
with exactly $N$ electrons when $\mu_N<E_F<\mu_{N+1}$. Charging
lines that are detected in the experiment correspond to $\mu_N=E_F$.

\subsection{Numerical procedure and self-consistency}
The Poisson equation \eqref{poisson} is solved on a three
dimensional mesh with a finite difference method. The adopted mesh
steps $\Delta x=\Delta y=12.5$ nm and $\Delta z=2$ nm are sufficient
to describe the charge distributions and the shapes of electrodes.
Smaller step in the growth direction is necessary because of a
strong localization of the electron gas at the interface. Same step
sizes are applied in the Schroedinger equation.

According to expression (7) the charge density that enters the
Poisson equation \eqref{poisson} is a sum of charge densities of
ionized donors that depend on the potential in a manner defined by
Eq. (8) and the electron gas accumulated at the GaAs/AlGaAs
interface. The latter depends on the total potential according to
Eq. (20). Electrons trapped by the quantum dot are the third charge
density [Eq.(26)] present in the nanodevice. The
Schroedinger-Poisson calculations are iterated till the
self-consistency of the three charge distributions with the total
potential is obtained. The iteration accounting for the dependence
of the electrons and ionized donors on the electrostatic potential
requires application of an under-relaxation technique to ensure
stability. The under-relaxed iteration is usually slowly convergent
and requires several hundred thousands iteration for the entire
mesh. The convergence is radically improved for a strategy of
simulated cooling of the system. The measurements are performed at
temperatures of several mK, for which occupation of electron states
in Eqs. (8) and (20) has a nearly binary distribution. For non-zero
temperatures we replace the formulas for the distribution of ionized
donors and the electron gas used in the above theory by expressions
accounting for the Fermi statistics
\begin{equation}
\rho_\mathrm{d}({\bf r})=\frac{|e|n_D({\bf
r})}{1+\exp\left(\frac{|e|\phi_\mathrm{tot}({\bf
r})+E^d}{kT}\right)}
\end{equation}
which tends to distribution given by (8) in $T=0$ limit. For the
electron gas at the interface we use the formula

\begin{equation}
\rho_\mathrm{d}({\bf r})=\frac{\tau
\rho^\mathrm{as}_\mathrm{el}(z)}{1+\exp\left(\frac{-|e|\phi_\mathrm{tot}(x,y,z_0)+U_b(z_0)}{kT}\right)},
\end{equation}
which in the limit $T\rightarrow 0$ tends to Eq. (20) with $\tau$
given in (21). We start the iteration at $T=15$ K, in which the
convergence is quickly reached. Next the temperature is gradually
reduced to the nominal value. With the simulated cooling the
convergence is obtained with an overall number of iterations that is
reduced ten times.

\section{Results}

The numerical procedure described above was applied for the device
described in Ref. [\cite{9}]. Basic parameters of the nanodevice
including composition and width of the semiconductor layers, the
position and shape of the gates, and the applied voltages are
adopted according to the experimental data.\cite{9} The theory
contains a single free parameter: the dopant concentration in the
AlGaAs barrier, which is not known with a sufficient precision since
the position of the transmission lines turns out to be extremely
sensitive to its even small variation. The donor density appearing
in formulas (8) and (28) was set to reproduce the charging of the
dot with the first electron when the voltages are set as in the
experiment (see below).

\begin{figure}[ht!]
\centerline{ \hbox{\epsfxsize=85mm
                \epsfbox[87 300 414 800] {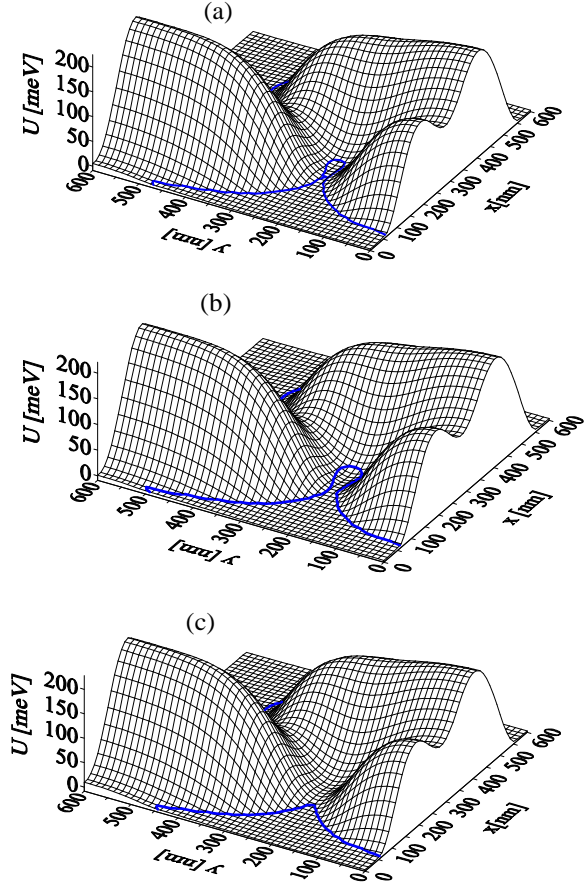}\hfill}}
\caption{Electron potential energy
$U(x,y)=-|e|\phi_\mathrm{elst}(x,y,z_o)$ calculated at a distance of
$z_o=12$ nm of the AlGaAs barrier for voltages $V_P=0$, $V_T=1.5$ V,
$V_M=-1.07$ V and $V_R=-0.096$ V. Assumed donor impurity
concentration is  $n_D=21.641 \times 10^{16}$ cm$^{-3}$ in (a),
 $n_D=25 \times 10^{16}$ cm$^{-3}$ in (b) and
  $n_D=20 \times 10^{16}$ cm$^{-3}$ in (c). Blue line shows the
 $U=0$ contour.
 \label{r4}}
\end{figure}

The two-dimensional electron gas in the asymptotic region acts as an
electron reservoir whose electrochemical potential is set by the
voltage of the source and drain that are connected to the electron
gas in the asymptotic region. We adopt the potential applied to the
source, drain and the electron gas as the reference value for the
voltages $V_S=V_D=0$. Following the experiment \cite{9} we assume
that voltages applied to the electrodes are: $V_P=0$, $V_T=1.5$ V,
$V_M=-1.07$ V and $V_R=-0.096$ V. Under these voltages a first
electron occupies the dot.\cite{9}  Figs. 4(c) and 4(b) present the
electrostatic potential distribution in $x,y$ plane for $z=z_o=12$
nm [used in Eq. (25)] calculated for $n_D=20\times10^{16}$cm$^{-3}$
and $n_D=25\times 10^{16}$cm$^{-3}$, respectively. In both figures
the blue line shows the zero of the electrostatic potential. Note,
that in Fig. 4(c) the zero potential is found far from the center of
the device which indicates that the dot cannot trap any electrons.
In Fig. 4(b) the zero level encircles quite a large region which
turns out to trap several electrons and not a single one. We found
that a single electron occupies the dot with the zero binding energy
for the donor concentration $n_D=21.641 \times 10^{16}$ cm$^{-3}$.
The corresponding potential profile in presented on a surface plot
in Fig. 4(a) and as a contour plot in Fig. 5(a). The shaded region
in Fig. 5(a) shows the electrodes. The potential has a well
developed minimum of negative potential that is just enough to trap
a single electron. The electron reservoir which encircles the dot is
presented in Fig. 6 which shows a density map of the electron gas
confined at the heterojunction. For the applied voltages the
oscillating plunger gate ($P$) voltage takes the electron out of the
dot to the reservoir and attracts it back with the electron
tunneling along the transmission channel that is opened parallel to
the $x$ direction. The opened channel is visible in Fig. 5(a) for
$x,y\simeq 200$ nm. In Fig. 5(b) and (c) the potential cavity is
surrounded by potential barriers that are too deep for the electron
to cross.

\begin{figure}[ht!]
\centerline{ \hbox{\epsfxsize=65mm
                \epsfbox[189 165 389 689] {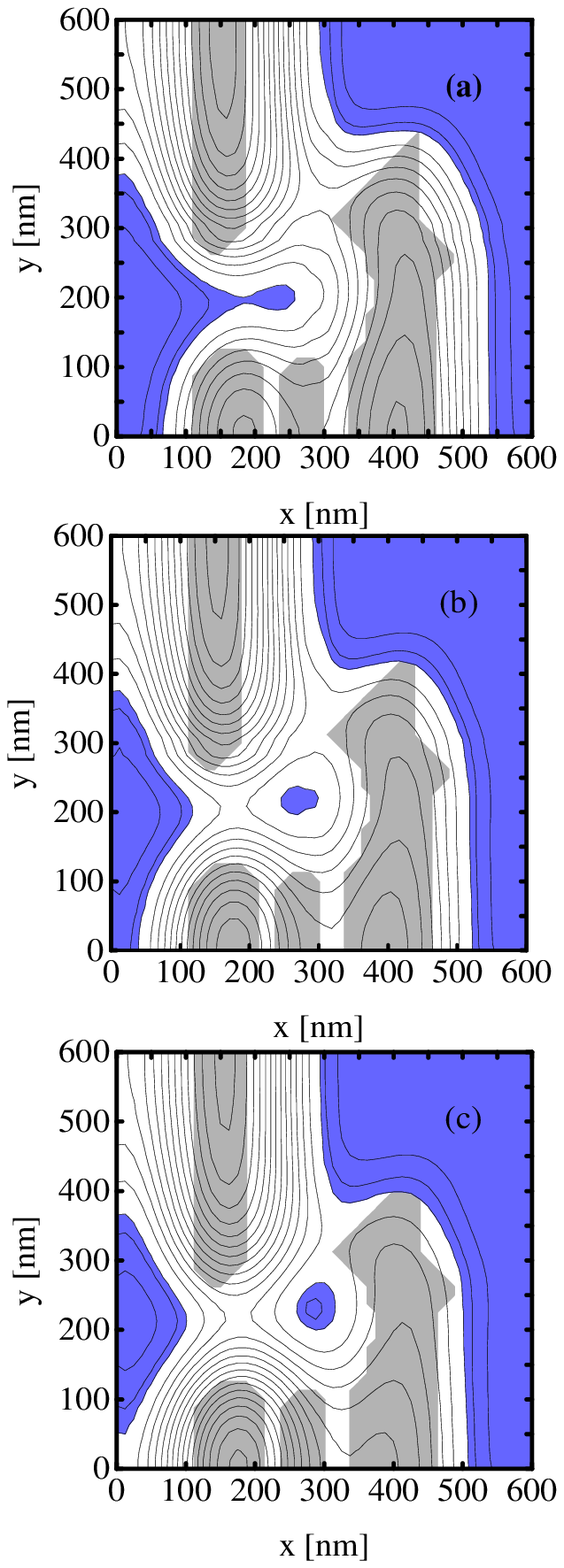}\hfill}}
\caption{(Color online) Equipotential contours and regions of
negative electron potential energy (plotted in blue) for
 $n_D=21.641 \times 10^{16}$ cm$^{-3}$ at 12 nm of the barrier.
 In (a) the voltages are same as in Fig. 4. In (b) and (c)
  (-1.12V,-0.93V)
and (-1.18V,-0.90V), respectively. Shaded regions show the metal
gates.
 \label{r5}}
\end{figure}

\begin{figure}[ht!]
\centerline{ \hbox{\epsfxsize=65mm
                \epsfbox[128 276 705 821] {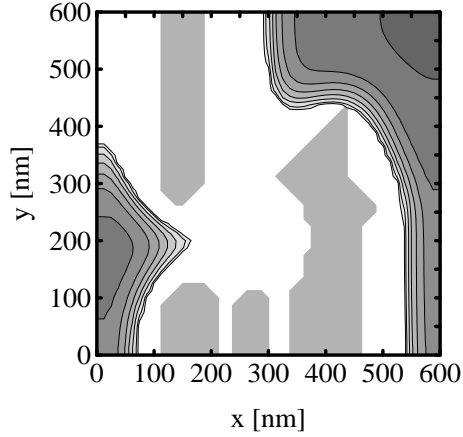}\hfill}}
\caption{Contour plot of the density of the electron gas at 12 nm of
the barrier for the voltages used in Fig. 6(a), the darker shade of
grey the larger density.}
 \label{genst}
\end{figure}

\begin{figure}[ht!]
\centerline{ \hbox{\epsfxsize=65mm
                \epsfbox[128 276 705 821] {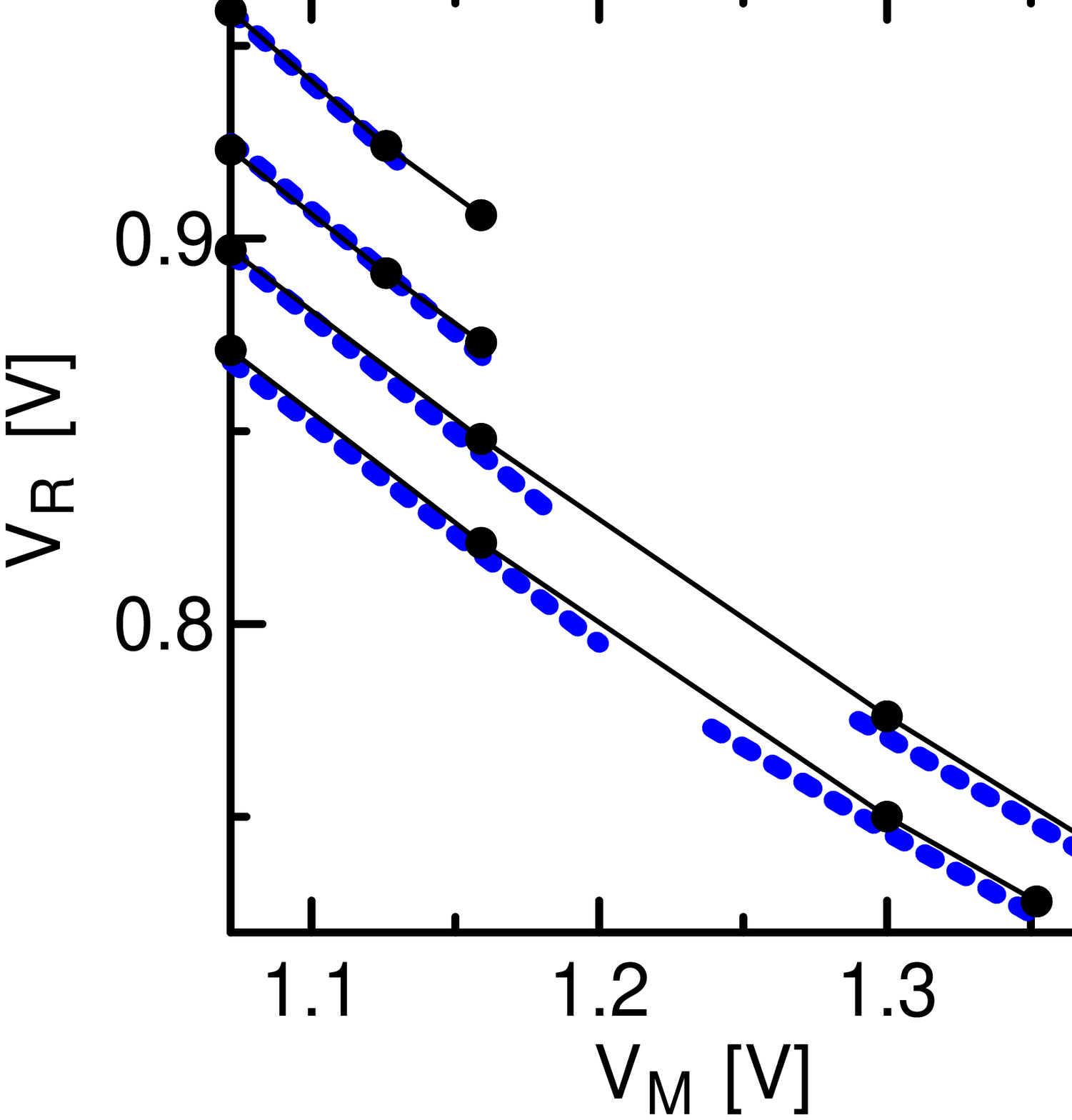}\hfill}}
\caption{Transmission lines as calculated (solid lines) and
measured\cite{9} (thick dashed lines). Above the highest line the
dot is empty.  }
\end{figure}

Formation of a single-electron artificial atom is obtained also for
other voltages. Lowered $V_M$ voltage results in an increase of the
electron potential energy in the center of the nanodevice that can
be compensated by an increase of $V_R$. A transmission line can be
traced on the $(V_M,V_R)$ plane. Similar transmission lines are
observed for the tunneling of the second electron from and into the
dot. In Fig. 7 we present the transmission lines calculated for 1-4
electrons which are compared to the experimental results.\cite{9} We
note that the calculations very well reproduce both the slope and
the positions of the transmission lines. The experimental lines for
the first and the second electron vanish for larger $V_M$, which is
due to the increase of the potential barrier closing the
transmission channel. For third and fourth electron the lines
reappear for larger $V_M$, which is explained by opening of a second
transmission channel. Increase of the potential barrier and
formation of the second transmission channel can be observed in
Figs. 5(b) and 5(c) which present the potential distribution with a
single dot-confined electron calculated for $V_M,V_R$ voltages equal
to (-1.12V,-0.93V) and (-1.18V,-0.90V), respectively. Fig. 5(b)
shows the case where the left channel of Fig. 5(a) is closed. In
Fig. 5(c) we see lowering of the potential barrier at  right of the
dot center. The tunneling probability is still not large enough to
allow for an experimental observation of the transmission.

Interaction of the confined electrons with the electron gas leads to
an appearance of nonlinear effects in the Schroedinger equation
\cite{25,26} for the dot-confined electrons. The non-linear effects
increase the electron binding energies and result in self-focusing
mechanism for a single confined electron. An account of this
self-focusing potential is taken in the Poisson equation
(\ref{poisson}). The Poisson equation does not account for the
energy relaxation of the dot-confined-electron system to the
reservoir.\cite{26} The energy relaxation leads to a decoherence of
the dot-confined quantum states. The energy relaxation rate is
likely to be significant due to small density of the electron gas in
the neighborhood of the dot.\cite{26}

\section{Summary}

We presented a theory describing phenomena appearing in a
multielectrode device of gated two-dimensional electron gas
containing a quantum dot. In the asymptotic region of a large
distance from the gated area the calculation consists in solution of
a single-dimensional Poisson-Schroedinger problem for the electron
gas in a strong magnetic field parallel to the semiconductor
surface. Solution in the asymptotic region is used to determine the
electron gas density in  the three-dimensional Poisson-Schroedinger
problem for the gated region with a quantum dot. The energies and
charge densities of several confined electrons were calculated with
a configuration interaction approach. The presented theory includes
a single fitting parameter - the donor concentration.  All the other
device parameters: the layer structure, the shape, size and position
of the electrodes as well as the applied magnetic field vector are
taken of the experiment. \cite{8,9,10} The slope and positions of
the transmission lines were reproduced with a very good quantitative
agreement with the experiment. We discussed the electric field
distribution in the device for voltages corresponding to
transmission lines observed in Ref. [9] with a particular attention
to the inhomogeneities creating the quantum dot confinement and the
surrounding potential barriers that separate the artificial atom of
the electron gas.

 For voltages corresponding to the transmission lines the weakest bound electron
 is
stimulated to oscillate between the dot and the reservoir by the
oscillating plunger voltage. We demonstrated opening and closing of
two transmission channels in the barriers that allow for the
oscillations of the confined charge.

\acknowledgments This work was
supported by the State Committee for Scientific Research  (KBN)
under Grant No. 1P03B 002 27.


\begin{thebibliography}{00}
\bibitem{1} D.P. DiVincenzo, Phys. Rev. A {\bf 51}, 1015 (1995).
\bibitem{2} D. Loss and D.P. DiVincenzo, Phys. Rev. A {\bf 57}, 120 (1998).
\bibitem{3} R.C. Ashoori, H.L. Stormer, J.S. Weiner, L.N. Pfeiffer, S.J. Pearton, K.W. Baldwin, and K.W. West, Phys. Rev. Lett. {\bf 68}, 3088
(1992). \bibitem{4} S. Tarucha, D.G. Austing, T. Honda, R.J. van der
Hage, and L.P. Kouwenhoven, Phys. Rev. Lett. {\bf 77}, 3613 (1996).
\bibitem{5} R.C. Ashoori, N.B. Zhitenev, L.N. Pfeiffer, and K.W. West,
Physica E {\bf 3}, 15  (1998). \bibitem{6} L.P. Kouwenhoven, T.H.
Osterkamp, M.W.S. Danoesastro, M. Eto, D.G. Austing, T. Honda, and
S. Tarucha, Science {\bf 278}, 1788 (1997). \bibitem{7} S. De
Franceschi, S. Sasaki, J.M. Elzerman, W.G. van der Wiel, S. Tarucha,
and L.P. Kouwenhoven, Phys. Rev. Lett. {\bf 86}, 878 (2001).
\bibitem{Molinari} M. Rontani, C. Cavazzoni, D. Bellucci, and G. Goldoni, J. Chem. Phys. {\bf 124}, 124102
(2006).
\bibitem{8} R. Hanson, B. Witkamp, L.M.K. Vandersypen, L.H. Willems van
Beveren, J.M. Elzerman, and L.P. Kouwenhoven, Phys. Rev. Lett. {\bf
91}, 196802 (2003). \bibitem{9} J.M. Elzerman, R. Hanson, L.H.
Willems van Beveren, L.M.K. Vandersypen, and L.P. Kouwenhoven, Appl.
Phys. Lett. {\bf 84}, 4617 (2004). \bibitem{10} J.M. Elzerman, R.
Hanson, L.H.Willems van Beveren, B. Witkamp, L.M.K. Vandersypen, and
L.P. Kouwenhoven, Nature {\bf 430}, 431 (2004). \bibitem{11} R.
Hanson, L.H. Willems van Beveren, I.T. Vink, J.M. Elzerman, W.J.M.
Naber, F.H.L. Koppens, L.P. Kouwenhoven, and L.M.K. Vandersypen,
Phys. Rev. Lett. {\bf 94}, 196802 (2005). \bibitem{12} T. Meunier,
I.T. Vink, L.H. Willems van Beveren, F.H.L. Koppens, H.P. Tranitz,
W. Wegscheider, L.P. Kouwenhoven, and L.M.K. Vandersypen, Phys. Rev.
B {\bf 74}, 195303 (2006). \bibitem{13}  J.R. Petta, A.C. Johnson,
J.M. Tylor, E.A. Laird, A. Yacoby, M.D. Lukin, C.M. Marcus, M.P.
Hanson, and A.C. Grossard, Science {\bf 309}, 2180 (2005).
\bibitem{14} J.M. Elzerman, R. Hanson, J.S. Greidanus, L.H.Willems
van Beveren, S. De Franceschi, L.M.K. Vandersypen, S. Tarucha and
L.P. Kouwenhoven, Physica E {\bf 25}, 135 (2004). \bibitem{15} J.
Kim, D.V. Melnikov, J.P. Leburton, D.G. Austing, and S. Tarucha,
Phys. Rev. B {\bf 74}, 035307 (2006).
\bibitem{kondo} W.G. van der Wiel, S. De Franceschi, T. Fujisawa, J.M. Elzerman, S. Tarucha, and L.P. Kouwenhoven,
Science {\bf 289}, 2105 (2000).
\bibitem{spindt} R. Hanson, L.H. Willems van Beveren, I.T. Vink,
J.M. Elzerman, W.J.M. Naber, F.H.L. Koppens, L.P. Kouwenhoven, and
L.M.K. Vandersypen, Phys. Rev. Lett. {\bf 94}, 196802 (2005).
\bibitem{spinrl} R. Hanson, B. Witkamp, L.M.K. Vandersypen, L.H. Willems van Beveren, J.M. Elzerman, and L.P.
Kouwenhoven, Phys. Rev. Lett. {\bf 91}, 196802 (2003).
\bibitem{16} N.A. Bruce and P.A. Maksym, Phys. Rev. B {\bf
61}, 4718 (2000). \bibitem{17} S. Bednarek, B. Szafran, and J.
Adamowski, Phys. Rev. B {\bf 61}, 4461 (2000). \bibitem{18} S.
Bednarek, B. Szafran, and J. Adamowski, Phys. Rev. B {\bf 64},
195303 (2001). \bibitem{19} B. Szafran, S. Bednarek, and J.
Adamowski, Phys. Rev. B {\bf 67}, 115323 (2003). \bibitem{20} P.
Matagne, J.P. Leburton, D.G. Austing, and S. Tarucha, Phys. Rev. B
{\bf 65}, 085325 (2002). \bibitem{21} P. Matagne and J.P. Leburton,
Phys. Rev. B {\bf 65}, 235323 (2002). \bibitem{22} S. Bednarek, B.
Szafran, K. Lis, and J. Adamowski, Phys. Rev. B {\bf 68}, 155333
(2003).
\bibitem{23} D.V. Melnikov, P. Matagne, J.P. Leburton, D.G.
Austing, G. Yu, S. Tarucha, J. Fettig, and N. Sobh, Phys. Rev. B
{\bf 72}, 085331 (2005). \bibitem{24} L.X. Zhang, P. Matagne, J.P.
Leburton, R. Hanson, and L.P. Kouwenhoven, Phys. Rev. B {\bf 69},
245301 (2004).
\bibitem{u} A. Weichselbaum and S.E. Ulloa, Phys. Rev. B {\bf 74},
085318 (2006).
 \bibitem{25} S. Bednarek, B. Szafran, and K. Lis,
Phys. Rev. B {\bf 72}, 075319 (2005). \bibitem{26} S. Bednarek and
B. Szafran, Phys. Rev. B {\bf 73}, 155318 (2006). \bibitem{uwa2} In
Ref. \cite{23} calculations for a rectangular vertical dot were
performed.
\bibitem{uwaga} In the limit of small mesh spacings $(\Delta x,\Delta y,\Delta z)$ the value of
$z_0$ has no influence on the energy of confined system, which can
be verified for the single-electron case. In the few-electron
calculations the spacings cannot be made very small. For a single
electron the results obtained with $z_0$ adopted variationally are
consistent with the zero mesh spacing limit.
\end{thebibliography}
\end{document}